\documentclass[fleqn,usenatbib]{mnras}
\usepackage{graphicx}
\usepackage{bm}
\usepackage{amssymb}
\usepackage{amsmath}
\usepackage{units}
\usepackage{array}
\usepackage{subcaption}
\usepackage{lmodern}
\usepackage{mathrsfs}

\title{Systematic Error in Approximate Models of the GRB Early Afterglow}

\author[B. Amend, E. R. Coughlin, and J. Zrake]{
Benjamin Amend,$^{1,2,3}$\thanks{E-mail: bamend@coastal.edu}
Eric R. Coughlin,$^{1}$
Jonathan Zrake$^{3}$
\\
$^{1}$Department of Physics, Syracuse University, Syracuse, NY 13210, USA\\
$^{2}$Department of Physics and Astronomy, Coastal Carolina University, Conway, SC 29528, USA\\
$^{3}$Department of Physics and Astronomy, Clemson University, Clemson, SC 29634, USA\\
}

\begin{document}
\maketitle

\begin{abstract}
Gamma-ray burst (GRB) afterglows are thought to arise when relativistic ejecta launched by a compact central engine drive a blast wave into the surrounding circumburst medium, producing broadband synchrotron emission. We present a rigorous assessment, based on high-resolution special relativistic hydrodynamics simulations, of a widely adopted `two-zone model' for approximating the dynamics of the early afterglow phase. Before the onset of the Blandford-McKee (BMK) self-similar solution, the outflow generally produces two emission components, associated with the forward-shocked circumburst medium and the reverse-shocked ejecta. The subsequent evolution depends on whether the reverse shock significantly decelerates the ejecta as it crosses the shell, separating the so-called relativistic and Newtonian reverse shock regimes. We show that when the reverse shock is Newtonian, it crosses the ejecta shell long before BMK self-similarity is established, leaving a prolonged interval that can span $\sim$ hours in observer time in which the true hydrodynamic evolution is not captured by standard semi-analytic prescriptions. We demonstrate that this mismatch can, for representative afterglow parameters, substantially overpredict the reverse-shock emission from radio through ultraviolet frequencies, or overpredict the forward-shock emission at X-ray frequencies, depending on how the transition away from the two-zone model is prescribed.
\end{abstract}


% =============================================================================
% Uncomment to discover the column and text width in inches:
% \makeatletter
% \def\convertto#1#2{\strip@pt\dimexpr #2*65536/\number\dimexpr 1#1}
% \makeatother
% Text width: \convertto{in}{\the\textwidth}, Column width: \convertto{in}{\the\columnwidth}.

\section{Introduction}
\label{sec:introduction}

Gamma-ray burst (GRB) afterglows are produced by synchrotron emission arising from the collision of ultra-relativistic ejecta with a circumburst medium \citep{1992MNRAS.258P..41R, 1993ApJ...418L...5P, 1997ApJ...476..232M, 1997ApJ...485L...5W, 1997MNRAS.288L..51W, 1998ApJ...497L..17S}. This interaction drives a forward shock, which sweeps up and heats the external gas, and a reverse shock that propagates back into the ejecta \citep{1995ApJ...455L.143S, 1997ApJ...476..232M}. The forward shock powers the long-lived, canonical afterglow, typically dominating the emission from hours to weeks after the burst across the electromagnetic spectrum, from the radio through X-ray bands \citep{1997ApJ...476..232M, 1997Natur.387..783C, 1997Natur.386..686V, 1997Natur.389..261F, 1998ApJ...497L..17S}. The reverse shock operates on much shorter timescales, and may produce optical flashes or radio flares in the minutes to hours following the prompt emission \citep{1997ApJ...476..232M, 1999ApJ...517L.109S, 1999ApJ...520..641S, 2000ApJ...545..807K, 1999Natur.398..400A, 1999ApJ...522L..97K}. These early-afterglow features arising from the reverse shock contain unique information about the ejecta and the central engine that powers the outflow \citep[e.g.][]{1999ApJ...517L.109S, 1999ApJ...520..641S, 2000ApJ...545..807K, 2003ApJ...582L..75K, 2004MNRAS.353..511P, 2013ApJ...776..119L, 2016ApJ...833...88L, 2017ApJ...848...69A, 2023NatAs...7..986B}.

Inferring the physical properties of GRB ejecta and their circumburst environments from afterglow observations requires a model for both the underlying shock dynamics and the resulting synchrotron emission. The latter depends on the densities, pressures, and bulk Lorentz factors of the forward- and reverse-shocked material, together with assumptions about the microphysics of particle acceleration and magnetic-field amplification. Relativistic hydrodynamic simulations can self-consistently provide the dynamical quantities entering the emission calculation, but are generally too computationally expensive to evaluate repeatedly within parameter-estimation procedures such as MCMC, which may require thousands of model realizations. Consequently, early-afterglow studies have traditionally relied on analytic or semi-analytic prescriptions for the forward- and reverse-shock dynamics \citep[e.g.][]{2000ApJ...545..807K,2003ApJ...582L..75K,2004MNRAS.353..511P,2005MNRAS.363...93Z}.

Several publicly available afterglow-modeling frameworks illustrate the range of dynamical approximations employed in practice. The \texttt{afterglowpy} package \citep{2020ApJ...896..166R} efficiently computes forward-shock emission from angularly structured jets, but does not explicitly evolve the reverse-shocked ejecta. \texttt{PyBlastAfterglow} includes both forward- and reverse-shock emission within a thin-shell treatment and adopts a pressure-balance formulation motivated by \citet{2013MNRAS.433.2107N}, making its dynamics closely related to the class of approximations considered here \citep{2025MNRAS.538.2089N}. Other recent frameworks have adopted more detailed treatments of the forward--reverse-shock system: the extension of \texttt{jetsimpy} to radially structured ejecta uses an energy-conservation prescription rather than pressure balance \citep{Wang_2025}, while \texttt{VegasAfterglow} similarly replaces the classical pressure-balance treatment with an energy-conserving prescription during reverse-shock crossing and separately evolves the shocked ejecta after crossing \citep{WANG2026100490}. \texttt{Magglow} additionally models a finite-duration transition phase that persists beyond reverse-shock crossing, during which the shocked ejecta continue to influence the forward shock before the flow approaches the Blandford-McKee regime \citep{10.1093/mnras/staf1923}. Despite the introduction of these more sophisticated treatments of the early blast-wave dynamics, simplified analytic and semi-analytic forward--reverse-shock prescriptions remain in active use for interpreting observed GRB afterglows and inferring their physical parameters \citep[e.g.][]{Chen_2024,10.1093/mnras/stad3243,Wu_2025,10.1093/mnras/staf1521,chastain2026grb240205breverseshock}. Such treatments often retain thin-/thick-shell classifications, analytic shock-crossing conditions, and prescribed post-crossing scalings \citep[e.g.][]{2000ApJ...545..807K,2005ApJ...628..315Z}, while pressure-balance thin-shell dynamics are also implemented in current general-purpose frameworks such as \texttt{PyBlastAfterglow} \citep{2025MNRAS.538.2089N}.

In one common class of treatments, the shocked material, termed the ‘blast’, is approximated as two uniform zones: the forward-shocked circumburst medium and the reverse-shocked ejecta \citep{1995ApJ...455L.143S, 2000ApJ...545..807K, 2005MNRAS.363...93Z}. The fluid variables in each zone are determined by solving the two-shock Riemann problem in planar geometry. At sufficiently late times, once the ejecta have transferred their energy to the external medium, the blast relaxes into the Blandford-McKee self-similar solution \citep{1976PhFl...19.1130B}, which provides the canonical description of the forward-shock afterglow. The late-time nature of the reverse-shocked ejecta is more ambiguous, and depends strongly on how relativistic the reverse shock is. If the reverse shock is relativistic, simulations show that the time-evolution of the volume-averaged fluid primitives in the shocked ejecta are accurately described by Blandford-McKee, with a density jump at the contact discontinuity \citep{2000ApJ...542..819K}. However, if the reverse shock is mildly relativistic or deeply Newtonian, the ejecta energy remains primarily kinetic, and as such cannot be described by the Blandford-McKee solution. In these cases, a common and well-motivated prescription is to assume $\gamma \propto R^{-g}$, where $\gamma$ is the volume-averaged fluid Lorentz factor of the shocked ejecta, $R$ is the radius of the forward shock, and $g$ is a dimensionless parameter constrained by the dynamics of the shocked circumburst medium \citep{1999MNRAS.306L..39M, 2000ApJ...542..819K}.

All of these late-time behavior prescriptions assume self-similarity in at least the forward-shocked region, and are thus not formally applicable prior to the time $t_{\mathrm{dec}}$ at which the state relaxes into a self-similar one. Meanwhile, the two-zone approximation itself has substantially degraded in accuracy by the time $t_{\Delta}$ at which the reverse shock has completely crossed the ejecta, occurring much earlier than relaxation into self-similarity when the reverse shock is Newtonian due to incomplete energy thermalization \citep{1995ApJ...455L.143S, 2000ApJ...545..807K, 2004MNRAS.353..647N}. Numerical simulations have likewise shown that the reverse-shock crossing does not correspond to an immediate transition to self-similarity; in RMHD simulations, \cite{2009A&A...494..879M} observed a post-crossing phase before the blast relaxed to the Blandford-McKee solution. This issue is further muddled by the conflation of $t_{\Delta}$ and $t_{\mathrm{dec}}$ in the literature, even though they are only comparable in the regime where the reverse shock is relativistic \citep{2005MNRAS.363...93Z, 2007ApJ...668.1083G, 2013ApJ...776..119L}. In the Newtonian reverse shock regime, the disparity between these times leads to a broad interval of evolution that is not well-described by either the two-zone approximation \textit{or} self-similarity, and we show in this work that this interval can span up to $\sim$ hours in observer time.

This poorly-modeled hydrodynamic phase can be masked in afterglow light curve fits by tuning microphysical parameters separately in the forward- and reverse-shocked regions. This can compensate for energy excesses or deficits in either region depending on the choice of transition timescale. Thus, even if a robust fit is achieved, the underlying hydrodynamics may still be misrepresented, and the inferred burst parameters may be sensitive to how the model handles the interval between $t_{\Delta}$ and $t_{\mathrm{dec}}$.

In this work, we assess the observable consequences of this uncertain post-crossing evolution by comparing prescriptions that switch away from the two-zone model at either $t_{\Delta}$ or $t_{\mathrm{dec}}$ with one-dimensional special relativistic hydrodynamic (SRHD) simulations. We compare the synchrotron spectra and light curves predicted by these prescriptions to those obtained from our simulations, with an emphasis on cases where the reverse shock is Newtonian. Our analysis shows that in this regime, either prematurely enforcing self-similarity or artificially prolonging the two-zone framework can misrepresent the dynamics, leading to systematic discrepancies in the predicted emission that may result in degeneracies in estimated burst parameters.

This paper is structured as follows. In Sec.~\ref{sec:grb_hydrodynamic_models}, we outline semi-analytic prescriptions for forward- and reverse-shock dynamics, the relevant timescales, and the hybrid transition models considered in this work. We introduce the code used for our one-dimensional SRHD simulations in Sec.~\ref{sec:numerical_simulations}, and cross-validate our simulations with the derived timescales. In Sec.~\ref{sec:consequences_for_observable_emission}, we outline our post-processing framework for the calculation of synthetic spectra and light curves, and compare these observables between the semi-analytic prescriptions and simulations. Finally, we discuss the implications of our findings in Sec.~\ref{sec:discussion}, and conclude in Sec.~\ref{sec:conclusions} with a summary of our results.

\section{GRB Hydrodynamic Models}
\label{sec:grb_hydrodynamic_models}
The hydrodynamic evolution of a GRB outflow can be divided into the following distinct stages:

\begin{enumerate}
    \item \textbf{Acceleration:} A baryon-loaded fireball converts its thermal energy into bulk motion, accelerating outward to relativistic speeds.
    \item \textbf{Coasting:} Once most of the thermal energy is depleted, the flow forms a cold, thin shell that coasts with an approximately constant Lorentz factor.
    \item \textbf{Interaction Regime:} Interaction with the ambient medium drives a forward shock into the external gas and a reverse shock into the ejecta, creating the shocked region (``blast") between them.
    Energy is continually transferred from the shell to the blast at least until the reverse shock fully traverses the ejecta.
    \item \textbf{Self-Similarity:} When the swept-up external mass carries energy comparable to that of the original ejecta, the system approaches the Blandford–McKee self-similar solution, where the forward shock decelerates and the blast evolves in a predictable, scale-free way.
    \item \textbf{Non-Relativistic Phase:} At much later times the blast transitions to Sedov-Taylor-like expansion (assuming the forward shock was not already sub-relativistic from the start), and eventually the forward shock dissipates into the ambient medium.
\end{enumerate}

We focus on stages (iii) and (iv) in this work, which encompass the reverse-shock-dominated interaction and the transition toward canonical self-similar evolution. %We use 
In the following subsections we %to 
outline the semi-analytic prescriptions that are commonly adopted to characterize %the behavior during 
these stages.

\subsection{The Early Afterglow: Interaction Regime}
\label{subsec:two_zone_model}
We consider a cold, ultra-relativistic shell of ejecta with lab-frame width $\Delta_{\mathrm{ej}}$, launched at radius $r_{\mathrm{ej}}$ and time $t_{\mathrm{ej}}=r_{\mathrm{ej}}/c$. The ejecta move with a bulk Lorentz factor $\gamma_4 = \gamma_{\mathrm{ej}} \gg 1$ into an external medium, the rest-mass density of which follows the power-law profile
\begin{equation}
    \rho_1(r) = \rho_{\mathrm{cb}}\left( \frac{r}{r_{\mathrm{ej}}} \right)^{-k}\,,
\end{equation}
where $k=0$ corresponds to a uniform circumburst medium and $k=2$ to a stellar wind-like environment, and $\rho_{\mathrm{cb}}$ is the normalization of the circumburst medium density. Because the shell is a spherical annulus that expands radially outward, the comoving density of the unshocked ejecta is
\begin{equation}
    \rho_4(r) = \rho_{\mathrm{ej}}\left( \frac{r}{r_{\mathrm{ej}}} \right)^{-2}\,,
\end{equation}
and the overdensity ratio $f \equiv \rho_4 / \rho_1$ is
\begin{equation}
    f(r) = f_{\mathrm{ej}}\left( \frac{r}{r_{\mathrm{ej}}} \right)^{k-2}\,,
\end{equation}
where $f_{\mathrm{ej}} = \rho_4(r_{\mathrm{ej}}) / \rho_1(r_{\mathrm{ej}}) = \rho_{\mathrm{ej}}/\rho_{\mathrm{cb}}$.

\begin{figure}
\begin{center}
\includegraphics{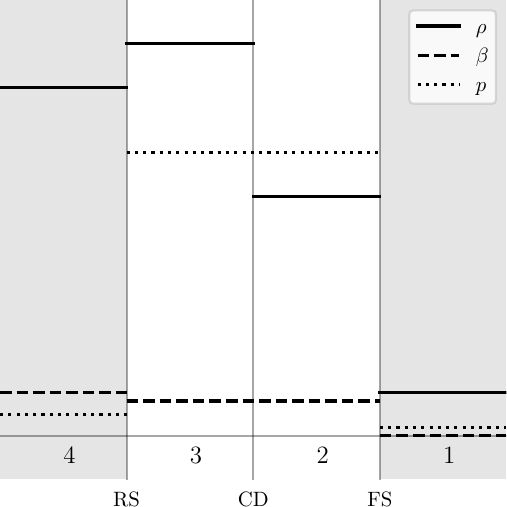}
\caption{
A schematic diagram of the fluid primitives in the `two-zone' model. The zones are: (4) unshocked ejecta, (3) shocked ejecta, (2) shocked circumburst medium, and (1) unshocked circumburst medium. The fluid primitives are spatially uniform within each zone, and the fluid states in zones 1 and 4 are known time-dependent boundary conditions.}
\label{fig:two_zone_model}
\end{center}
\end{figure}

As the shell interacts with the ambient gas, %a pair of 
two shocks form: a forward shock propagating into the circumburst medium and a reverse shock propagating back into the ejecta. The structure of the flow during the interaction stage is often subdivided into four uniform regions (Fig.~\ref{fig:two_zone_model}): (1) the unshocked circumburst medium, (2) the shocked circumburst medium, (3) the shocked ejecta, and (4) the unshocked ejecta, where regions 2 and 3 comprise the ``blast.''

The fluid in each zone is characterized by sets of conserved quantities $U$ and fluxes $F$,
\begin{equation}
    U = \begin{pmatrix}\gamma\rho\\\gamma^2\beta\rho h\\ \gamma^2\rho h - \gamma\rho - p \end{pmatrix}\,,
    \hspace*{0.5cm}
    F = \begin{pmatrix}\gamma\beta\rho\\\gamma^2\beta^2\rho h + p\\ \gamma^2\beta\rho h - \gamma\beta\rho \end{pmatrix}\,.
    \label{eq:conserved_quantities}
\end{equation}
where $\beta = v/c$, $\gamma=(1-\beta^2)^{-1/2}$, $\rho$ is the comoving mass density, $p$ is the pressure, and $h$ is the specific enthalpy. Using a gamma-law equation of state, the specific enthalpy can be expressed as
\begin{equation}
    h = 1 + \frac{p\hat{\gamma}}{\rho(\hat{\gamma}-1)}\,,
    \label{eq:enthalpy_density}
\end{equation}
where $\hat{\gamma}$ is the adiabatic index, taken to be $4/3$ for relativistic flow. The changes in these conserved quantities across a shock are related to the lab-frame shock speed $\beta_{\mathrm{shock}}$ component-wise via
\begin{equation}
    \beta_{\mathrm{shock}} = \frac{\Delta F_j}{\Delta U_j}\,,
\end{equation}
yielding the Rankine-Hugoniot jump conditions,
\begin{equation}
    \begin{aligned}
        \beta_{\mathrm{shock}} [\gamma\rho] = [\gamma\beta\rho] \\
        \beta_{\mathrm{shock}} [\gamma^2\beta\rho h] = [\gamma^2\beta^2\rho h + p] \\
        \beta_{\mathrm{shock}} [\gamma^2\rho h - \gamma\rho - p] = [\gamma^2\beta\rho h - \gamma\beta\rho]
        \label{eq:jump_conditions}
    \end{aligned}
\end{equation}
where the square brackets indicate a difference between the right and left states relative to the forward or reverse shock.

Generally, this system of equations does not yield a closed-form solution without additional approximations, and several procedural approaches have been developed to make the problem more tractable. One strategy is to derive simplified solutions in particular asymptotic limits, \citep[e.g.][]{2005MNRAS.363...93Z}. A more general approach is to construct fitting formulae that reproduce these solutions, \citep[e.g.][]{2022MNRAS.513.4887Z}. Another common simplification is to collapse the two-zone structure into a single zone, treating the entire shocked region as a uniform, infinitesimally thin shell at the forward shock front \citep[e.g.][]{2024ApJS..273...17W}, though this particular approximation only makes sense when the reverse shock contributes negligibly to the dynamics. More recently, \citet{Wang_2025} extended this framework to include the reverse shock by treating the forward- and reverse-shocked regions as separate thin shells, while replacing pressure balance with a global energy-conservation prescription.

Nevertheless, the most general method involves applying these jump conditions locally at each shock surface, assuming spatial uniformity within zones 2 and 3 and treating zones 1 and 4 as time-dependent boundary conditions \citep{1995ApJ...455L.143S, 2004RvMP...76.1143P, 2005MNRAS.363...93Z, 2013MNRAS.433.2107N}. Assuming that $\beta_2=\beta_3$ and $p_2=p_3$ to preserve the nature of the contact discontinuity, Eqs.~\ref{eq:jump_conditions} and \ref{eq:enthalpy_density} constitute a closed set of equations that characterizes the evolution of zones 2 and 3. This system can be reduced to a single equation for the common $\gamma\beta$ between zones 2 and 3 which can be solved via numerical root-finding techniques; all other fluid primitives in zones 2 and 3 then follow analytically. We refer to this prescription as the `two-zone model' going forward, since only zones 2 and 3 are evolved dynamically, and zones 1 and 4 are treated as time-dependent boundary conditions.

This two-zone model provides an accurate description of the hydrodynamics of a spherical blast in planar geometry, and it remains valid over short times in spherical geometry. As the blast expands, the decreasing ejecta density and evolving shock conditions produce gradients in the density and pressure profiles over $\sim$ a few dynamical times. Once these gradients become appreciable across the shocked regions, the assumption that each region can be represented by a single uniform state begins to break down, and the resulting internal structure is no longer well-captured by the two-zone model. Continuing to enforce the two-zone pressure-equilibrium prescription beyond this point can artificially redistribute thermal energy between the shocked regions, particularly after the reverse shock has crossed the ejecta and no longer supplies newly shocked material to region 3. This issue was noted in \cite{2006ApJ...651L...1B}, where a `mechanical model' was developed to try to resolve this issue. While this mechanical model provides a more physically robust framework for modeling the hydrodynamics of the blast, closely related pressure-balance thin-shell formulations remain implemented in current afterglow-modeling frameworks such as \texttt{PyBlastAfterglow} \citep{2025MNRAS.538.2089N}.

\subsection{The Late Afterglow: Self-Similarity}
\label{subsec:blandford_mckee_self_similar_solution}

The late afterglow is hydrodynamically simpler than the early afterglow since emission from the reverse shock quickly diminishes once the energy supply of unshocked ejecta is depleted, leaving predominantly emission from the forward shock, which begins to relax into the Blandford-McKee (BMK) self-similar solution \citep{1976PhFl...19.1130B} for an adiabatic impulsive blast wave. This solution arises under the assumptions of spherical symmetry, negligible radiative losses, and a strong shock propagating with Lorentz factor $\Gamma \gg 1$. It is characterized by self-similar scaling relations for the fluid variables behind the forward shock:
\begin{equation}
    p = \frac{2}{3}w_1\Gamma^2F(\chi)\,, \hspace{0.5cm} \gamma^2 = \frac{1}{2}\Gamma^2G(\chi)\,, \hspace{0.5cm} \rho' = 2\rho_1 \Gamma^2 H(\chi)\,,
\end{equation}
where $\gamma$ is the local Lorentz factor of the fluid, $\rho'$ is the lab-frame mass density in the shocked fluid, $w_1$ is the enthalpy of the unshocked ambient medium, and $\rho_1$ is its rest-mass density. The ambient density declines as $r^{-k}$, and the functions $F(\chi)$, $G(\chi)$, and $H(\chi)$ describe the radial structure of the blast in terms of the self-similar variable
\begin{equation}
    \chi = \left[1 + 2(m+1)\Gamma^2\right](1-r/t)\,,
\end{equation}
where the scaling parameter $m$ relates to the external density profile via $m = (3 - k) > -1$ in the ultra-relativistic limit, and $\chi=1$ corresponds to the location of the forward shock.

The BMK solution offers an excellent description of the hydrodynamics of the blast at late times \citep{1999ApJ...513..669K, 2009ApJ...698.1261Z}, when reverse shock contributions are negligible. However, emission from the shocked ejecta is not expected to cease abruptly once the reverse shock has fully traversed the shell, but rather taper over time. When the reverse shock is relativistic, this tapering is characterized by the same self-similar behavior as the forward-shocked region, with a density jump at the contact. For a Newtonian reverse shock, the bulk Lorentz factor assumes a more general power-law form, where
\begin{equation}
    \gamma_{3} \propto R_{\mathrm{CD}}^{-g}\,,
    \label{eq:post-crossing-1}
\end{equation}
where $R_{\mathrm{CD}}$ is the radius of the contact discontinuity, and $(3-k)/2 \leq g \leq (7-2k)/2$ \citep{1999MNRAS.306L..39M, 2000ApJ...542..819K}. Analogous scaling relationships can be obtained for the other fluid primitives by assuming the heated ejecta expand adiabatically, leading to
\begin{equation}
    \rho_3 \propto R_{\mathrm{CD}}^{-6(3+g)/7}\,, \hspace{0.5cm} p_3 \propto R_{\mathrm{CD}}^{-8(3+g)/7}\,.
    \label{eq:post-crossing-2}
\end{equation}
These scalings approximate the behavior of the averaged fluid primitives across the entire ejecta shell. Note that the purpose of this work is not to test the validity of these scalings; \cite{2000ApJ...542..819K} already found excellent agreement with representative Newtonian and relativistic reverse shock simulations. Rather, we highlight these prescriptions here because they provide the post-crossing evolution adopted for the reverse-shocked ejecta in the hybrid transition models defined in Sec.~\ref{subsec:hybrid-prescription}.

\subsection{Transition from Early to Late Afterglow}
\label{subsec:timescales}

Two dynamical timescales characterize the global evolution of the blast \citep{1995ApJ...455L.143S}. The reverse-shock crossing time, $t_{\Delta}$, marks the instant when the reverse shock has completely traversed the ejecta shell. The deceleration time, $t_{\mathrm{dec}}$, occurs when the bulk of the ejecta kinetic energy has been transferred to the shocked circumburst medium and the system begins to approach the Blandford-McKee self-similar regime. The ordering and relative magnitude of these times depend on how relativistic the reverse shock is, which is governed by the instantaneous ratio $f/\gamma^2$, where $\gamma$ is the common Lorentz factor of the two shocked regions. When $f/\gamma^2 \gg 1$, the reverse shock is Newtonian, processing the ejecta inefficiently such that $t_{\Delta} \ll t_{\mathrm{dec}}$. When $f/\gamma^2 \ll 1$, the reverse shock is relativistic and efficiently thermalizes the flow such that $t_{\Delta} \sim t_{\mathrm{dec}}$.

This distinction relates to the standard thin- and thick-shell classification in GRB hydrodynamic literature \citep[e.g.][]{1995ApJ...455L.143S, 1999ApJ...513..669K, 2000ApJ...542..819K}. These designations refer to the shell width relative to the deceleration length scale: in a `thin' shell the reverse shock does not become relativistic prior to deceleration, whereas it can in a `thick' shell. The thinness/thickness of the shell is considered relative to the deceleration length scale $r_{\mathrm{dec}}$, such that if $\Delta_{\mathrm{ej}} \ll r_{\mathrm{dec}}/\gamma_{\mathrm{ej}}^2$, the reverse shock crosses the ejecta before the blast has undergone substantial global deceleration, consistent with a Newtonian reverse shock. Conversely, when $\Delta_{\mathrm{ej}} \sim r_{\mathrm{dec}}/\gamma_{\mathrm{ej}}^2$, the reverse shock has sufficient time to become relativistic before crossing the shell, and the crossing and deceleration times become comparable.

\subsubsection{Newtonian Reverse Shock}
\label{ssub:nrs}

In the Newtonian reverse shock (NRS) regime, the reverse shock converts only a small fraction of the shell's bulk kinetic energy into thermal energy as it traverses the ejecta. The subsequent deceleration of the blast is then mediated by the inward pressure gradient force developing in the forward-shocked region, and the ejecta are considered decelerated once the energy of the swept-up external mass becomes comparable to the initial kinetic energy of the shell. The total energy of the ejecta is
\begin{equation}
    E_{\mathrm{shell}} = \gamma_{\mathrm{ej}}M_{\mathrm{ej}}c^2\,.
\end{equation}
The energy imparted to the shocked circumburst medium can be approximated as $(4/3)\pi r_{\mathrm{dec}}^3\rho_1\Gamma_{12}^2c^2$, and equating these two energies yields the standard deceleration condition \citep{1992MNRAS.258P..41R}
\begin{equation}
    E_{\mathrm{shell}} \simeq \frac{4\pi}{3}\rho_1\Gamma_{12}^2r_{\mathrm{dec}}^3c^2\,.
\end{equation}
More generally, for a power-law circumburst medium the integrated swept-up energy becomes
\begin{equation}
    E_{\mathrm{shell}} \simeq 4\pi\rho_{\mathrm{cb}}\Gamma_{12}^2\frac{r_{\mathrm{dec}}^3c^2}{3-k}\left( \frac{r_{\mathrm{ej}}}{r_{\mathrm{dec}}} \right)^k\,,
\end{equation}
assuming $r_{\mathrm{dec}} \gg r_{\mathrm{ej}}$ and $k < 3$. Substituting $E_{\mathrm{shell}} = 4\pi r_{\mathrm{ej}}^2\Delta_{\mathrm{ej}}\gamma_{\mathrm{ej}}^2\rho_{\mathrm{cb}}f_{\mathrm{ej}}c^2$ and using $\Gamma_{12} \sim \gamma_{\mathrm{ej}}$ (since the reverse shock only impacts the ejecta by a negligible amount) yields the characteristic deceleration timescale in the NRS regime,
\begin{equation}
    t_{\mathrm{dec,NRS}} \sim t_{\mathrm{ej}} \left[ (3-k)f_{\mathrm{ej}}\left( \frac{\Delta_{\mathrm{ej}}}{r_{\mathrm{ej}}} \right) \right]^{1/(3-k)}\,.
    \label{eq:t_dec_nrs}
\end{equation}

This timescale characterizes the bulk conversion of kinetic energy from the shell to thermal energy in the swept-up circumburst medium, marking the onset of relaxation into self-similarity.

\subsubsection{Relativistic Reverse Shock}
\label{ssub:rrs}

In the regime where $f/\gamma^2 \ll 1$, the reverse shock becomes relativistic in the rest frame of the unshocked ejecta. In this relativistic reverse shock (RRS) regime, the shock efficiently decelerates the cold ejecta, and the bulk of the energy transfer from the shell to the circumburst medium occurs over shorter timescales.

The key difference from the NRS regime is that the reverse shock now efficiently decelerates the ejecta, so the shocked fluid Lorentz factor must be determined self-consistently from pressure balance rather than approximated as $\gamma_{12} \sim \gamma_{\mathrm{ej}}$. The post-shock pressure in each region may be expressed using the relativistic jump conditions,
\begin{equation}
    p_d = (\gamma_{ud} - 1)(\hat{\gamma}\gamma_{ud} + 1)\rho_u\,,
\end{equation}
where subscripts $u$ and $d$ refer to the upstream and downstream fluids, respectively, $\gamma_{ud}$ is the Lorentz factor of the upstream measured in the downstream frame, and $\hat{\gamma}$ is the adiabatic index of the shocked gas. Pressure equilibrium across the contact discontinuity requires $p_2=p_3$; in the RRS limit, $\gamma_{43}\simeq \gamma_{\mathrm{ej}}/\gamma_{12} \gg 1$, and this condition gives
\begin{equation}
    \gamma_{12}^2\rho_1 \simeq \left( \frac{\gamma_{\mathrm{ej}}}{\gamma_{12}} \right)^2\rho_4\,.
\end{equation}
Expressing this in terms of the overdensity ratio $f(r)$ yields
\begin{equation}
    \gamma_{12}^4 \simeq \gamma_{\mathrm{ej}}^2f_{\mathrm{ej}}\left( \frac{r}{r_{\mathrm{ej}}} \right)^{k-2}\,.
\end{equation}
This relation allows the RRS deceleration radius (time) to be expressed directly in terms of the shell and ambient parameters. Substituting this into the condition that the swept-up energy equals the ejecta energy gives
\begin{equation}
    t_{\mathrm{dec,RRS}} \sim t_{\mathrm{ej}} \left[ (3-k)\gamma_{\mathrm{ej}}\sqrt{f_{\mathrm{ej}}}\left( \frac{\Delta_{\mathrm{ej}}}{r_{\mathrm{ej}}} \right) \right]^{2/(4-k)}\,.
    \label{eq:t_dec_rrs}
\end{equation}
The Lorentz factor of the shocked fluid at $t_{\mathrm{dec}}$ can be estimated from the same scaling as
\begin{equation}
    \gamma_{\mathrm{dec,RRS}} \sim \gamma_{\mathrm{ej}}^{2/(4-k)}f_{\mathrm{ej}}^{(k-2)/(8-2k)}\left( \frac{\Delta_{\mathrm{ej}}}{r_{\mathrm{ej}}} \right)^{(k-2)/(4-k)}\,,
    \label{eq:gamma_dec_rrs}
\end{equation}
which should be interpreted as the initial bulk Lorentz factor at the onset of Blandford-McKee (and comparable to the forward shock Lorentz factor to within a factor of $\sim \sqrt{2}$). For a uniform external medium ($k=0$), the deceleration time reduces to $t_{\mathrm{dec,RRS}} \propto t_{\mathrm{ej}}\gamma_{\mathrm{ej}}^{1/2}f_{\mathrm{ej}}^{1/4}(\Delta_{\mathrm{ej}}/r_{\mathrm{ej}})^{1/2}$, identical to the reverse-shock crossing time reported by \cite{1995ApJ...455L.143S} (equivalent to their $R_{\Delta}/c$). This equivalence between $t_{\Delta}$ and $t_{\mathrm{dec}}$ is to be expected in the RRS regime, since the reverse shock is highly efficient at thermalizing the kinetic energy in the cold ejecta.

\subsubsection{Reverse Shock Crossing Time}
\label{ssub:rs_cross}

The reverse shock crossing time $t_{\Delta}$ can be estimated by integrating the velocity difference between the unshocked ejecta and the reverse shock over the duration of the crossing,
\begin{equation}
    \Delta_{\mathrm{ej}} = \displaystyle\int_{t_{\mathrm{ej}}}^{t_{\Delta}}[\beta_{\mathrm{ej}} - \beta_{\mathrm{RS}}(t)]\,dt\,,
\end{equation}
where all quantities are measured in the lab frame.

The lab-frame speed of the reverse shock, $\beta_{\mathrm{RS}}$, is approximately related to the relative speed between the shocked and unshocked ejecta by the standard velocity addition formula,
\begin{equation}
    \beta_{\mathrm{RS}} \simeq \frac{\beta_{\mathrm{ej}} - \beta_{43}}{1 - \beta_{\mathrm{ej}}\beta_{43}}\,,
\end{equation}
where $\beta_{43}$ is the velocity of the shocked ejecta as measured in the unshocked ejecta frame. To leading order, $\beta_{43}$ can be determined from the pressure-balance condition across the contact discontinuity, which for arbitrary reverse shock strength gives
\begin{equation}
    \gamma_{12}^2 \simeq f(\gamma_{43}^2 - 1)\,.
\end{equation}
Expressing $\gamma_{43}$ in terms of the known quantities $\gamma_{\mathrm{ej}}$ and $f$ yields
\begin{equation}
    \gamma_{43} \simeq \left[ 1 + \frac{1 - 2\sqrt{\gamma_{\mathrm{ej}}^2/f}}{f/\gamma_{\mathrm{ej}}^2 - 4} \right]\,,
\end{equation}
which leads to the approximate scaling
\begin{equation}
    \beta_{\mathrm{ej}} - \beta_{\mathrm{RS}} \simeq \frac{1}{\gamma_{\mathrm{ej}}\sqrt{f}}\,.
\end{equation}
Integrating this velocity difference and equating the result to the initial shell width gives the crossing time,
\begin{equation}
    t_{\Delta} \sim t_{\mathrm{ej}}\left[ \left( \frac{4-k}{2} \right) \gamma_{\mathrm{ej}}\sqrt{f_{\mathrm{ej}}}\left( \frac{\Delta_{\mathrm{ej}}}{r_{\mathrm{ej}}} \right) \right]^{2/(4-k)}\,.
    \label{eq:rs_crossing_time}
\end{equation}
Again, for $k=0$, $t_{\mathrm{\Delta}} \sim t_{\mathrm{ej}}\gamma_{\mathrm{ej}}^{1/2}f_{\mathrm{ej}}^{1/4}(\Delta_{\mathrm{ej}}/r_{\mathrm{ej}})^{1/2}$, equivalent to $R_{\Delta}/c$ in \cite{1995ApJ...455L.143S}, who also highlighted the insensitivity of this result to how relativistic the reverse shock is. Indeed, Eq. \ref{eq:rs_crossing_time} holds regardless of whether the reverse shock is relativistic or Newtonian. Note that Eqs. \ref{eq:rs_crossing_time} and \ref{eq:t_dec_rrs} are equivalent to within the constant factor $[(4-k)/(6-2k)]^{2/(4-k)}$, which is of order unity for $0 \leq k < 2$, and identically unity for $k=2$. Both $t_{\Delta,\mathrm{RRS}}$ and $t_{\mathrm{dec,RRS}}$ are entirely different from Eq. \ref{eq:t_dec_nrs}, however, and this discrepancy is a reflection of the deceleration efficiency (or inefficiency) of the reverse shock.

\subsubsection{Additional Transitional Thresholds}
\label{ssub:additional_transitional_thresholds}

A useful timescale is the time $t_{\mathrm{NR}}$ at which the forward shock becomes non-relativistic,
\begin{equation}
    t_{\mathrm{NR}} = t_{\mathrm{ej}} \left[ \frac{17-4k}{2}\gamma_{\mathrm{ej}}^2f_{\mathrm{ej}} \left( \frac{\Delta_{\mathrm{ej}}}{r_{\mathrm{ej}}} \right) \right]^{1/(3-k)}\,.
\end{equation}
This timescale comes from equating the initial energy of the ejecta shell to the energy of the Blandford-McKee solution when the forward shock Lorentz factor is $\sim 1$. In this work, we generally restrict our analysis to ultra-relativistic outflows for which $t_{\mathrm{NR}} \gg t_{\mathrm{dec}}$, and the blast remains highly relativistic throughout the timescales of interest.

The reverse shock may start out Newtonian but become relativistic as the blast evolves. To determine when the reverse shock transitions from Newtonian to relativistic, we introduce a critical Lorentz factor $\gamma_{\mathrm{crit}}$ defined by the condition $f/\gamma^2=1$. A relationship can be derived between $\gamma_{\mathrm{crit}}$ and the Sedov length $L_{\mathrm{sed}}$,
\begin{equation}
    L_{\mathrm{sed}} \equiv r_{\mathrm{ej}} \left[ (3-k)f_{\mathrm{ej}}\gamma_{\mathrm{ej}}\left( \frac{\Delta_{\mathrm{ej}}}{r_{\mathrm{ej}}} \right) \right]^{1/(3-k)}\,,
\end{equation}
defined to be the radius at which the swept-up mass is equal to the ejecta mass $M_{\mathrm{ej}}$ (and assuming $L_{\mathrm{sed}} \gg r_{\mathrm{ej}}$). The overdensity ratio $f$ can be expressed in terms of $L_{\mathrm{sed}}$, and the condition $f/\gamma^2=1$ can be solved for $\gamma_{\mathrm{crit}}$ by letting $r=r_{\mathrm{dec,NRS}}$ and $\gamma_{\mathrm{ej}}=\gamma_{\mathrm{crit}}$, yielding the expression
\begin{equation}
    \gamma_{\mathrm{crit}} = \left[ \frac{1}{3-k} \left( \frac{L_{\mathrm{sed}}}{\Delta_{\mathrm{ej}}} \right) \right]^{(3-k)/(7-2k)}\,.
\label{eq:gamma_crit}
\end{equation}
Note that $k=2$ is a special case where $f(r)$ and $\gamma(r)$ are constant prior to $t_{\Delta}$. In this scenario, the criteria $f/\gamma^2=1$ and $f_{\mathrm{ej}}/\gamma_{\mathrm{ej}}^2=1$ are the same.

\subsection{Hybrid Transition Prescriptions}
\label{subsec:hybrid-prescription}

The two-zone model described in Section 2.1 and the late-time prescriptions described in Section 2.2 characterize limiting stages of the blast evolution, but do not uniquely specify how the system should transition between them. To examine the consequences of this ambiguity, we construct two hybrid prescriptions that differ only in the choice of a transition time, $t_{\rm tr}$. For $t<t_{\rm tr}$, both the forward- and reverse-shocked regions are evolved according to the two-zone model. For $t\geq t_{\rm tr}$, the forward-shocked circumburst medium is instead described by the Blandford-McKee solution, while the reverse-shocked ejecta evolve according to the post-crossing scalings of Eqs.~\ref{eq:post-crossing-1} and \ref{eq:post-crossing-2}.

The Blandford-McKee solution is normalized using the initial ejecta energy $E_{\rm shell}$, rather than rescaled to match the instantaneous forward-shocked energy at $t_{\rm tr}$. This ensures the correct asymptotic energy normalization once the ejecta energy has been transferred to the circumburst medium, but can introduce a discontinuity when BMK evolution is imposed before $t_{\rm dec}$.

The post-transition reverse-shocked ejecta are normalized to the two-zone state at $t_{\rm tr}$, such that
\begin{equation}
    \begin{aligned}
        \gamma_3=\gamma_{3,\rm tr}\left(\frac{R_{\rm CD}}{R_{\rm CD,tr}}\right)^{-g}\,, \\
        \rho_3=\rho_{3,\rm tr}\left(\frac{R_{\rm CD}}{R_{\rm CD,tr}}\right)^{-6(3+g)/7}\,, \\
        p_3=p_{3,\rm tr}\left(\frac{R_{\rm CD}}{R_{\rm CD,tr}}\right)^{-8(3+g)/7}\,,
    \end{aligned}
\end{equation}
where the subscript ``tr" denotes the corresponding two-zone quantity evaluated at $t_{\rm tr}$.

We consider two choices for $t_{\rm tr}$. In the first, $t_{\rm tr} = t_{\Delta}$, so the two-zone treatment is terminated when the reverse shock has completely crossed the ejecta; we refer to this prescription as `Two-Zone ($t_{\Delta}$)'. In the second, $t_{\rm tr}=t_{\rm dec}$; the two-zone treatment is retained until the forward shock begins to approach the Blandford-McKee regime, and we refer to this prescription as `Two-Zone ($t_{\rm dec}$)'. In the NRS regime, where $t_{\Delta} \ll t_{\rm dec}$, these choices bracket the uncertain transition interval. The former terminates the two-zone treatment when energy injection into the reverse-shocked ejecta ceases, but imposes Blandford-McKee evolution on the forward-shocked material before it has reached self-similarity. The latter delays the forward-shock transition until approximately the appropriate dynamical time, but continues to enforce pressure balance with the reverse-shocked ejecta after the reverse shock has crossed the shell. These hybrid prescriptions are intended as diagnostic limiting cases rather than as fully self-consistent descriptions of the intermediate evolution.

For the $k=2$ hybrid models considered in Secs.~\ref{subsec:two_zone_model_comparison} and \ref{subsec:transition_timescale_choices}, we adopt $g=1/2$, which qualitatively reproduces the scaling of the volume-averaged ejecta Lorentz factor with contact radius. We verified that varying $g$ over its physically allowed range, $1/2 \leq g \leq 3/2$, has only a minor effect on the post-processed spectra considered here.

\section{SRHD Simulations}
\label{sec:numerical_simulations}

\subsection{Numerical Methods}
\label{subsec:numerical_methods}

To assess the accuracy of the semi-analytic prescriptions outlined in prior sections, and to provide a numerical benchmark for comparisons of hydrodynamic behavior, we perform high-resolution 1D SRHD simulations of spherical blast waves.

\subsubsection{1D SRHD Code}
\label{ssub:1d_srhd_code}

As outlined in Eq. \ref{eq:conserved_quantities}, the conserved variables are the lab-frame mass density $D=\gamma\rho$, radial momentum density $S=\gamma^2\rho hv$, and energy density excluding rest mass $\tau = \gamma^2\rho h - \gamma\rho - p$, which together form the conserved state vector $U = (D, S, \tau)$ (note the code units are such that $c=1$). The system is closed with a gamma-law equation of state (EoS),
\begin{equation}
    p = (\hat{\gamma} - 1)\rho e \, ,
\end{equation}
where $\hat{\gamma}$ is the adiabatic index and $e$ is the specific internal energy. We assume a relativistic equation of state, such that $\hat{\gamma} = 4/3$.

We evolve the equations of SRHD in flux-conservative form \citep{2003LRR.....6....7M} in spherical symmetry, including the relevant geometric source terms,
\begin{equation}
    \frac{\partial D}{\partial t} + \frac{1}{r^2}\frac{\partial}{\partial r}\left( r^2Dv \right) = 0
\end{equation}
\begin{equation}
    \frac{\partial S}{\partial t} + \frac{1}{r^2}\frac{\partial}{\partial r} \left[ r^2(Sv + p) \right] = \frac{2p}{r}
    \label{eq:radial-momentum}
\end{equation}
\begin{equation}
    \frac{\partial \tau}{\partial t} + \frac{1}{r^2}\frac{\partial}{\partial r}\left[ r^2(\tau + p)v \right] = 0\,.
\end{equation}
Our code employs a moving-mesh finite-volume method to evolve the conserved variables. Fluxes $\hat{F}$ across cell interfaces are computed using the HLL approximate Riemann solver:
\begin{equation}
    \hat{F}^j_{i + 1/2} =
    \begin{cases}
    F_L^j - v_{i+1/2} U_L^j, & \text{if } v_{i+1/2} < a^- \\
    F_R^j - v_{i+1/2} U_R^j, & \text{if } v_{i+1/2} > a^+ \\
    \displaystyle F_{\mathrm{HLL}}^j
    - v_{i+1/2} U_{\mathrm{HLL}}^j \displaystyle, & \text{otherwise}\,.
    \end{cases}
\end{equation}
Here, $(L, R)$ refer to the fluid states immediately to the left or right of the cell interface at position $i + 1/2$, reconstructed to linear order using the standard minmod slope limiter. The interface speeds $v_{i+1/2}$ are chosen so that the domain expands homologously to follow the flow in the blast region, $v_{i+1/2} = r_{i + 1/2} / t$. The HLL flux is given by
\begin{equation}
    F_{\mathrm{HLL}}^j = \frac{a^+ F_L^j - a^- F_R^j + a^+ a^- (U_R^j - U_L^j)}{a^+ - a^-}\,,
\end{equation}
with the corresponding HLL-averaged conserved state given by
\begin{equation}
    U_{\mathrm{HLL}}^j = \frac{a^+ U_L^j - a^- U_R^j + F_L^j - F_R^j}{a^+ - a^-}\,.
\end{equation}
The minimum and maximum signal speeds, $a^-$ and $a^+$, are computed as
\begin{equation}
    a^- = \mathrm{min}\left( \frac{v_L - c_L}{1 - v_L \cdot c_L}, \frac{v_R - c_R}{1 - v_R \cdot c_R} \right)\,,
\end{equation}
\begin{equation}
    a^+ = \mathrm{max}\left( \frac{v_L + c_L}{1 + v_L \cdot c_L}, \frac{v_R + c_R}{1 + v_R \cdot c_R} \right)\,,
\end{equation}
where $c_L$ and $c_R$ are the sound speeds in the left and right states, respectively.

The code achieves second-order spatial accuracy using a piecewise-linear method with slope-limited reconstruction. Time integration is performed via a second-order Runge-Kutta scheme. Each Runge-Kutta sub-step is taken by advancing the volume-integrated vector of the $j$th conserved quantity in zone $i$ at time level $n$ according to
\begin{equation}
    \Delta U_i^j = \Delta t\left[-(\hat F^j_{i+1/2} \cdot A_{i+1/2} - \hat F^j_{i-1/2} \cdot A_{i-1/2}) + S_i^j \right]
\end{equation}
where $A_{i\pm1/2} = 4 \pi r_{i\pm1/2}^2$ is the area of the spherical interface at radius $r_{i\pm1/2}$, and $S_i^j$ is the volume-integrated geometric source term (Eq. \ref{eq:radial-momentum}) for the $j$th conserved quantity in zone $i$.

\subsubsection{Simulation Setup}
\label{ssub:simulation_setup}

We initialize ejecta fluid primitive variables $\mathbb{P}=(\rho, u, p)$ using the profile
\begin{equation}
    \mathbb{P}(r) = \frac{\mathbb{P}_0}{1 + e^{-\alpha(r-r_{\mathrm{ej}}+\Delta_{\mathrm{ej}})/r_{\mathrm{ej}}}}\,,
\end{equation}
where $\alpha$ is a smoothness parameter. This slightly smoothed profile prevents numerical artifacts associated with perfectly sharp edges while preserving the desired ejecta shell structure. The external medium is characterized by a stationary power-law density profile, $\rho(r) = \rho_{\mathrm{cb}}r^{-k}$.

We configure all simulations so that the ejecta launch time $t_{\mathrm{ej}}=r_{\mathrm{ej}}/c$, the deceleration time $t_{\mathrm{dec}}$, and the transition to non-relativistic evolution $t_{\mathrm{NR}}$ satisfy $t_{\mathrm{ej}} \ll t_{\mathrm{dec}} \ll t_{\mathrm{NR}}$. The blast region is resolved with $10^6-10^7$ zones depending on the particular configuration, which we verified through convergence tests to be sufficient for accurately capturing the relevant structure.

We conducted two sets of simulations: one where $k=0$, and another where $k=2$, initializing all simulations with a shell width $\Delta_{\mathrm{ej}} = 0.01 r_{\mathrm{ej}}$. Within each ambient-medium suite, we varied the initial Lorentz factor $\gamma_{\mathrm{ej}}$ while holding the ejecta mass $M_{\mathrm{ej}}$ fixed. Since
\begin{equation}
    M_{\mathrm{ej}} = 4\pi r_{\mathrm{ej}}^2
    \Delta_{\mathrm{ej}} f_{\mathrm{ej}}\rho_{\mathrm{cb}}\gamma_{\mathrm{ej}}\,,
\end{equation}
this requires $f_{\mathrm{ej}}\propto \gamma_{\mathrm{ej}}^{-1}$. The mass normalization was chosen separately for the two suites, corresponding to Sedov lengths $L_{\mathrm{sed}} = 10^2 r_{\mathrm{ej}}$ for $k=0$ and $L_{\mathrm{sed}} = 10^5 r_{\mathrm{ej}}$ for $k=2$. The Lorentz-factor ranges were then selected to span broadly across the dynamically relevant reverse-shock regimes. For $k=0$, these parameters give $\gamma_{\rm crit}\sim 32$, above which an initially Newtonian reverse shock becomes relativistic before crossing the shell, while $f_{\rm ej}/\gamma_{\rm ej}^2=1$ occurs at $\gamma_{\rm ej}\sim 320$. We therefore sample $3.16 \leq \gamma_{\rm ej} \leq 3160$, extending approximately one decade below the lower threshold and one decade above the upper threshold. For $k=2$, the two thresholds coincide at $\gamma_{\rm ej} \sim 2.1 \times 10^2$, and we correspondingly sample $21.2 \leq \gamma_{\rm ej} \leq 2120$, approximately one decade on either side. Both suites use the same logarithmic spacing, $\Delta\log_{10}(\gamma_{\rm ej}) = 1/15$, resulting in 46 simulations for $k=0$ and 31 for $k=2$.

\subsection{Timescales}
\label{subsec:timescale_comparison}

\begin{figure}
\begin{center}
\includegraphics{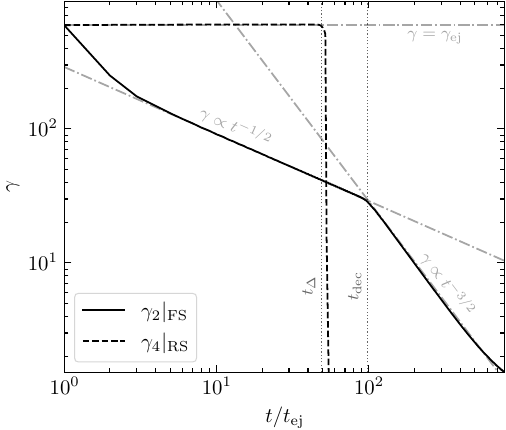}
\caption{
Time series of the post-forward-shocked circumburst medium ($\gamma_{2}|_{\mathrm{FS}}$) and the pre-reverse-shocked ejecta ($\gamma_{4}|_{\mathrm{RS}}$) for a $k=0$ blast wave in the RRS regime where $f_{\mathrm{ej}} = 5.928\times 10^4$ and $\gamma_{\mathrm{ej}}=5.623\times 10^2$. There is a precipitous drop in $\gamma_{4}|_{\mathrm{RS}}$ as the reverse shock finishes sweeping through the ejecta shell, and a broken power-law knee in $\gamma_{2}|_{\mathrm{FS}}$ as the post-forward-shocked fluid begins to transition from a $t^{-1/2}$ scaling (typical of an RRS blast wave such as this one) into the expected Blandford-McKee $t^{-3/2}$ scaling.}
\label{fig:time_measurement}
\end{center}
\end{figure}

The initial smoothing applied to the interior edge of the ejecta shell makes timescale measurements from the simulations somewhat nontrivial, particularly for $t_{\Delta}$. Nevertheless, there are distinct `knees' in the time series of both the pre-reverse-shocked ejecta at $t_{\Delta}$ and the post-forward-shocked circumburst medium at $t_{\mathrm{dec}}$. We measure the relevant timescales as the points of maximal curvature $\kappa$ in these time series, where
\begin{equation}
    \kappa = \frac{|\gamma''(t)|}{[1+\gamma'^2(t)]^{3/2}}\,.
\end{equation}
Here, $\gamma$ is the pre-reverse-shocked ejecta Lorentz factor (termed $\gamma_{4}|_{\mathrm{RS}}$) when solving for $t_{\Delta}$, and the post-forward-shocked circumburst medium Lorentz factor ($\gamma_{2}|_{\mathrm{FS}}$) for $t_{\mathrm{dec}}$ (Fig. \ref{fig:time_measurement}).

Both timescales can be output directly from the two-zone model, without simulations. The reverse shock crossing time is simply the time at which the position of the reverse shock equals the position of the inner edge of the ejecta shell. Similarly, we take the deceleration time to be the time at which the energy of the swept-up ambient material is equal to the initial ejecta energy.

\begin{figure*}
    \centering
    \begin{minipage}{0.48\textwidth}
        \centering
        \includegraphics[width=\linewidth]{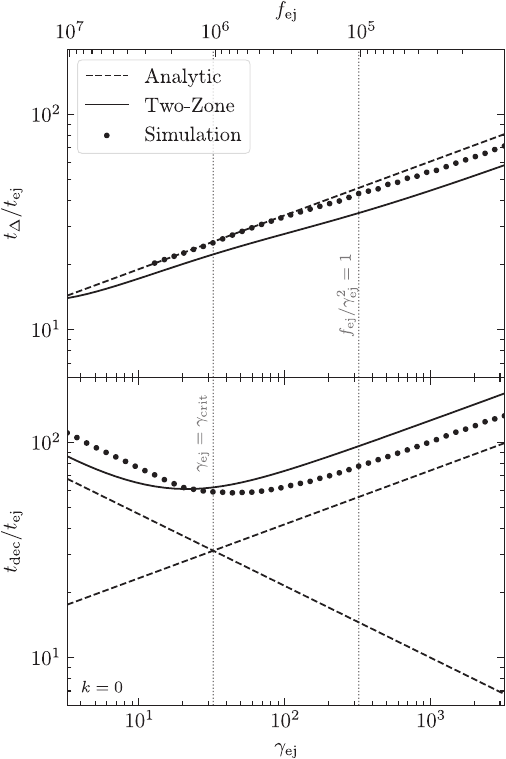}
    \end{minipage}
    \hfill
    \begin{minipage}{0.48\textwidth}
        \centering
        \includegraphics[width=\linewidth]{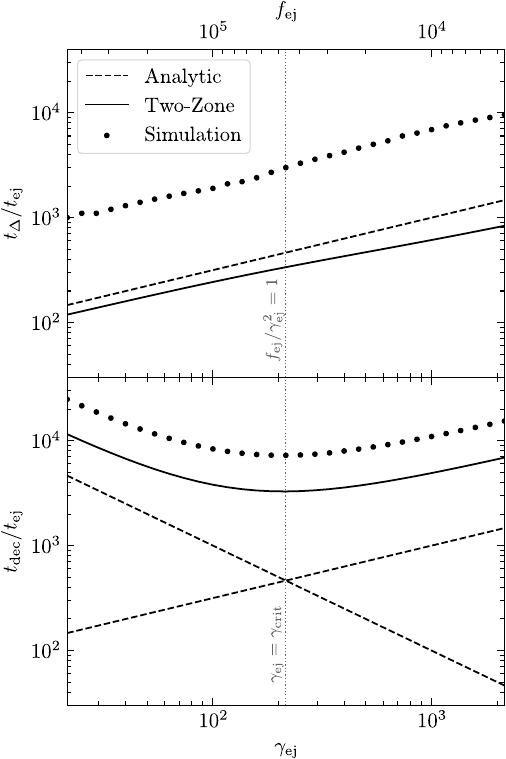}
    \end{minipage}
    \caption{Reverse shock crossing and deceleration timescales vs. ejecta Lorentz factors and overdensity parameters for a series of initial conditions that span from the NRS through RRS regimes in ISM ($k=0$, \textbf{left}) and wind-like ($k=2$, \textbf{right}) circumburst media. Both $\Delta_{\mathrm{ej}}=0.01r_{\mathrm{ej}}$ and $M_{\mathrm{ej}}=4\pi r_{\mathrm{ej}}^2\Delta_{\mathrm{ej}}f_{\mathrm{ej}}\rho_{\mathrm{cb}}\gamma_{\mathrm{ej}}$ were held fixed across all instances for each type of ambient medium. Shown are the analytic predictions from Sec.~\ref{subsec:timescales} as dashed lines, the two-zone model predictions (Sec. \ref{subsec:two_zone_model}) as solid lines, and our simulation results as scatter plot points. Also plotted are vertical lines that correspond to $\gamma_{\mathrm{crit}}$ (the critical Lorentz factor above which the reverse shock \textit{will become} relativistic before crossing the shell) and $f_{\mathrm{ej}}/\gamma_{\mathrm{ej}}^2=1$ (the threshold separating reverse shocks that \textit{start out} in the NRS or RRS regimes); note that these are equivalent in the external wind scenario. For the ISM case in particular, the two-zone model predicts a $\gamma_{\mathrm{crit}}$ that is lower than the analytic $\gamma_{\mathrm{crit}}$; this is because the ejecta shell is artificially inflated when applying the two-zone model at times $t > t_{\Delta}$, lengthening the lifespan of the reverse shock and allowing it to accelerate to relativistic speeds even at smaller $\gamma_{\mathrm{ej}}$. A few deep-NRS $k=0$ simulation measurements are omitted where $t_{\Delta}$ could not be robustly identified by the curvature-based diagnostic.}
    \label{fig:timescales}
\end{figure*}

Both the reverse shock crossing times and deceleration times measured from our simulations exhibit the same scaling behavior as the analytic predictions from Sec. \ref{subsec:timescales}, as seen in Fig. \ref{fig:timescales}. We also find that these same scaling relations are reproduced in the two-zone model as well - even $t_{\mathrm{dec}}$, despite no longer providing a physically self-consistent thermal-energy evolution for the reverse-shocked ejecta once the reverse shock has crossed the shell. Thus, we see independent lines of evidence for the behavior of $t_{\Delta}$ and $t_{\mathrm{dec}}$ across analytic (Sec. \ref{subsec:timescales}), semi-analytic (Sec. \ref{subsec:two_zone_model}), and numerical (Sec. \ref{subsec:numerical_methods}) methodologies, validating the disparate behaviors of these two timescales in the NRS regime.

\subsection{Shortcomings of the Two-Zone Model}
\label{subsec:two_zone_model_comparison}

\begin{figure}
\begin{center}
\includegraphics{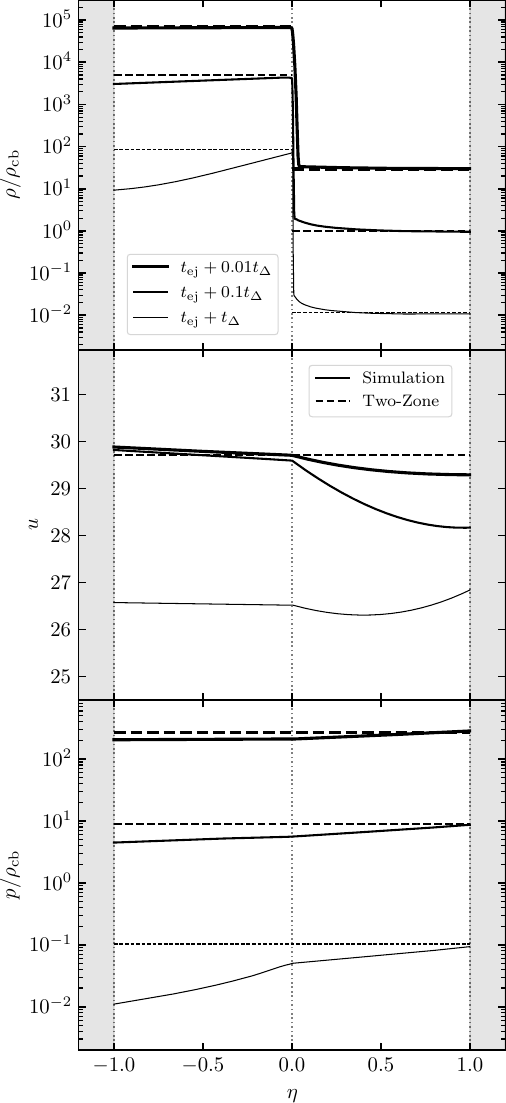}
\caption{A comparison of fluid primitive radial profiles as predicted by the two-zone model and as observed in simulations at various times up to and including $t_{\Delta}$ for a blast in a wind-like medium ($k=2$) in the NRS regime. The variable $\eta$ represents $(r - R_{\mathrm{CD}}) / (R_{\mathrm{CD}} - R_{\mathrm{RS}})$ for $R_{\mathrm{RS}} \leq r \leq R_{\mathrm{CD}}$, and $(r - R_{\mathrm{CD}})/(R_{\mathrm{FS}} - R_{\mathrm{CD}})$ for $R_{\mathrm{CD}} \leq r \leq R_{\mathrm{FS}}$. Expectedly, the simulation gradually drifts away from the two-zone model as the system evolves, and radial structure forms in both shocked regions.}
\label{fig:overlay}
\end{center}
\end{figure}

The two-zone model is only an accurate description of the hydrodynamics as long as the blast remains thin enough that pressure waves cross on timescales shorter than $r/c$. Once this is no longer the case, radial structure begins to develop, generally at times much earlier than $t_{\Delta}$; this can be clearly seen in Fig. \ref{fig:overlay}.

\begin{figure}
\begin{center}
\includegraphics{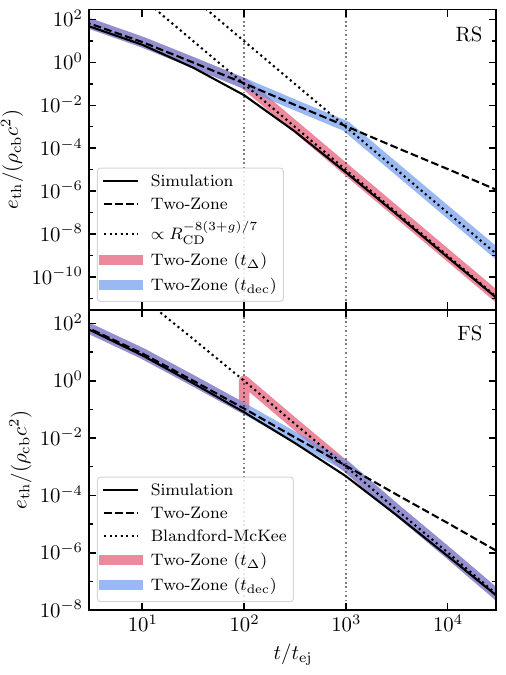}
\caption{Dimensionless thermal energy density in the reverse- (\textbf{top}) and forward- (\textbf{bottom}) shocked regions for a blast in a wind-like ($k=2$) environment with a Newtonian reverse shock. Two hybrid prescriptions defined in Sec.~\ref{subsec:hybrid-prescription} are shown: one where the switch to late-stage prescriptions occurs at $t_{\Delta}$ (`Two-Zone ($t_{\Delta}$)'), and one where the switch occurs at $t_{\mathrm{dec}}$ (`Two-Zone ($t_{\mathrm{dec}}$)'), with $g=1/2$ in both instances. The simulation data favors the earlier switch for the reverse-shocked region, and the later switch for the forward-shocked region. Selecting either of these timescales as a singular fiducial switching time either substantially overpredicts the thermal energy density in the reverse-shocked region indefinitely, or overpredicts it in the forward-shocked region during the transition window between $t_{\Delta}$ and $t_{\mathrm{dec}}$.}
\label{fig:energy_density}
\end{center}
\end{figure}

A more significant source of potential error lies in the choice of transition timescale when switching from the two-zone model to the self-similar or near-self-similar prescriptions that better characterize the late afterglow phase. An important physical input for synchrotron post-processing is the total energy; because the volumes are comparable between the two-zone output and our simulations, we compare the comoving thermal energy densities $e_{\mathrm{th}}$ directly (see Fig.\ref{fig:energy_density}). We find that the $t_{\rm tr} = t_{\Delta}$ hybrid prescription defined in Sec.~\ref{subsec:hybrid-prescription} matches the thermal energy density in the reverse-shocked region, but results in an excess of thermal energy in the forward-shocked material. One could in principle rescale the Blandford-McKee solution to align with the two-zone model at $t_{\Delta}$, but doing so would under-predict the thermal energy budget in the forward-shocked region, and would never converge to the simulation output. Conversely, the $t_{\rm tr} = t_{\mathrm{dec}}$ prescription instead provides a better fit for the forward-shocked energy content, but overpredicts $e_{\mathrm{th}}$ in the reverse-shocked ejecta. These differences are relatively inconsequential in the RRS regime since $t_{\Delta} \sim t_{\mathrm{dec}}$, but they are significant in the NRS regime where $t_{\Delta} \ll t_{\mathrm{dec}}$ (Fig. \ref{fig:energy_density}). This has substantial implications for observables, which we discuss in the following section.

\section{Spectra \& Light Curve Modeling}
\label{sec:consequences_for_observable_emission}

\subsection{Synchrotron Post-Processing}
\label{subsec:observables}

We synthesize spectra by post-processing each zone in our hydrodynamic simulations. We assume that a fraction $\epsilon_e$ of the post-shock internal energy is carried by non-thermal electrons with electron Lorentz factor $\gamma_e$ and a fraction $\epsilon_B$ is carried by magnetic fields. The comoving magnetic field strength is therefore
\begin{equation}
    B = \sqrt{8\pi\epsilon_Be_{\mathrm{th}}}\,,
\end{equation}
where $e_{\mathrm{th}}$ is the comoving thermal energy density. Electrons are assumed to be accelerated at the shocks into a power-law energy distribution,
\begin{equation}
    n(E)\,dE = KE^{-p}\,dE\,, \hspace{0.5cm} E \geq E_{\mathrm{min}} = \gamma_{\mathrm{min}}m_ec^2\,,
\end{equation}
where $p > 2$ and the normalization constant $K$ is fixed by the total electron number density and charge neutrality, $n_e \simeq \rho/m_p$. Equating the mean electron energy to $\epsilon_e e_{\mathrm{th}}/n_e$ gives
\begin{equation}
    \gamma_{\mathrm{min}} = \mathrm{max}\left[1,\,\epsilon_e\left( \frac{p-2}{p-1} \right)\frac{e_{\mathrm{th}}}{\rho c^2}\frac{m_p}{m_e}\right]\,,
\end{equation}
and
\begin{equation}
    K = \frac{\rho(p-1)E_{\mathrm{min}}^{p-1}}{m_p}\,.
    \label{eq:electron_normalization}
\end{equation}
Radiative losses introduce a cooling Lorentz factor, $\gamma_c$, defined by the condition that the synchrotron cooling time equals the comoving time $t'$ since shock heating \citep{1998ApJ...497L..17S}:
\begin{equation}
    \gamma_c = \frac{6\pi m_ec}{\sigma_TB^2t'}\,.
\end{equation}
If $\gamma_c > \gamma_{\mathrm{min}}$, then $E_c > E_{\mathrm{min}}$ and the zone is treated as `slow-cooling', where we use
\begin{equation}
    n(E) = \begin{cases} 
              KE^{-p}\,, & E_{\mathrm{min}} \leq E < E_c\,, \\
              KE_cE^{-(p+1)}\,, & E\geq E_c\,.
           \end{cases}
\end{equation}
If $\gamma_c < \gamma_{\mathrm{min}}$, $E_c < E_{\mathrm{min}}$ and the zone is treated as `fast-cooling', where we use
\begin{equation}
    n(E) = \begin{cases} 
              AE_{\mathrm{min}}^{1-p}E^{-2}\,, & E_c \leq E < E_{\mathrm{min}}\,, \\
              AE^{-(p+1)}\,, & E\geq E_{\mathrm{min}}\,,
           \end{cases}
\end{equation}
with
\begin{equation}
    A = \frac{\rho E_{\mathrm{min}}^p}{m_p}\left[ \frac{pE_c}{pE_{\mathrm{min}}+(1-p)E_c} \right]\,.
\end{equation}

The synchrotron power for a single electron is \citep{1979rpa..book.....R}
\begin{equation}
    P_{\nu}(\nu, \gamma_e) = \frac{\sqrt{3}}{2\pi}\frac{q_e^3B\sin{\alpha}}{m_ec^2}F\left( \frac{\nu}{\nu_c} \right)\,, \hspace{0.5cm} F(x) = x\displaystyle\int_x^{\infty}K_{5/3}(\eta)\,d\eta\,,
\end{equation}
where $\nu_c = (3/2)\gamma_e^2\nu_G\sin{\alpha}$ and $\nu_G = q_eB / (2\pi m_e c)$; we average over pitch angle $\alpha$. We then compute the comoving power per unit frequency from all electrons,
\begin{equation}
    P_{\nu,\mathrm{tot}}(\nu) = \displaystyle\int n(E)P_{\nu}(\nu,E)\,dE\,,
\end{equation}
and the spectral luminosity of a uniform zone of volume $V$ is $L_{\nu}' = P_{\nu,\mathrm{tot}}V$.

Transformation to the observer frame includes relativistic boosting and beaming. For an on-axis observer, the Doppler factor is $\mathscr{D}=[\gamma(1-\beta)]^{-1}$ (where $\gamma$ is the fluid Lorentz factor). Frequencies and fluxes transform as $\nu_{\mathrm{obs}} = \mathscr{D} \nu'$ and $F_{\nu,\mathrm{obs}}=\mathscr{D}^3F_{\nu}'/(4\gamma^2)$, where the factor $1/4\gamma^2$ approximates the solid-angle beaming of relativistic emission. The total observed flux is then obtained by summing over all emitting fluid elements in both the forward- and reverse-shocked regions,
\begin{equation}
    F_{\nu,\mathrm{obs}} = \displaystyle\sum_i \frac{\mathscr{D}_i^3L'_{\nu',i}}{16\pi\gamma_i^2 d^2}\,,
\end{equation}
where $d$ is the source distance.

In these diagnostic calculations, we intentionally adopt a simplified radiation treatment in order to isolate differences arising from the underlying hydrodynamics. We omit equal-arrival-time integration, angular structure across the visible emitting region, and synchrotron self-absorption, and assume fixed microphysical parameters $\epsilon_{B}$, $\epsilon_e$, and $p$. These approximations affect the detailed spectral and temporal appearance of the emission: equal-arrival-time and angular integration would smooth features produced over short dynamical intervals, while self-absorption can substantially modify the low-frequency spectrum. The adopted microphysical parameters and cooling prescription likewise determine the locations of the synchrotron spectral breaks and therefore which shocked region dominates at a particular observing frequency. Radiative losses are included in the synchrotron post-processing but are not coupled back to the hydrodynamic evolution, so our comparison also assumes that the blast remains approximately adiabatic over the timescales considered here. The spectra and light curves presented below should consequently be interpreted as controlled comparisons between hydrodynamic prescriptions rather than complete predictions for an individual GRB. We discuss the robustness of our conclusions to these assumptions further in Sec.~\ref{sec:discussion}.

\subsection{Spectra \& Light Curves}
\label{subsec:transition_timescale_choices}

To examine the observable consequences of the hydrodynamic discrepancies identified in earlier sections, we post-process a single representative simulation in the NRS regime. In this selected case, we take $f_{\mathrm{ej}}=10^5$, $\gamma_{\mathrm{ej}}=31.6$, and $\Delta_{\mathrm{ej}} = 0.01\,r_{\mathrm{ej}}$, with a wind-like ($k=2$) circumburst medium. For these parameters, $f_{\mathrm{ej}}/\gamma_{\mathrm{ej}}^2 \simeq 100$, so the reverse shock is deeply Newtonian and remains inefficient at thermalizing the ejecta as it crosses the shell. Consequently, the reverse-shock crossing and deceleration times are widely separated: $t_{\Delta}=100\,t_{\mathrm{ej}}$, while $t_{\mathrm{dec}}=10^3\,t_{\mathrm{ej}}$. The reverse shock therefore crosses the ejecta after only the first tenth of the evolution toward deceleration, after which the system is no longer necessarily well-described by either the two-zone model or the Blandford-McKee solution.

To dimensionalize this case, we assign the shell an isotropic-equivalent energy $E_{\mathrm{shell}} = 10^{53}\,\mathrm{erg}$ and fix the launch radius to $r_{\mathrm{ej}}=10^{14}\,\mathrm{cm}$. This gives a lab-frame time unit $t_{\mathrm{ej}} = r_{\mathrm{ej}}/c = 3.34\times 10^{3}\,\mathrm{s}$. The corresponding observer times are estimated using $t_{\mathrm{obs}}\simeq R_{\mathrm{CD}}/(2c\gamma^2)$, where $\gamma$ is taken to be the volume-averaged fluid Lorentz factor across the blast region. The reverse-shock crossing and deceleration times then occur at $t_{\Delta,\mathrm{obs}} \simeq 241\,\mathrm{s}$ and $t_{\mathrm{dec,obs}} \simeq 5.6 \times 10^3\,\mathrm{s}$, respectively. Thus, for this representative burst, the transition window between the two-zone and Blandford-McKee limits spans the first $\sim$1-2 hours of observer-frame spectral evolution.

\begin{figure}
\begin{center}
\includegraphics{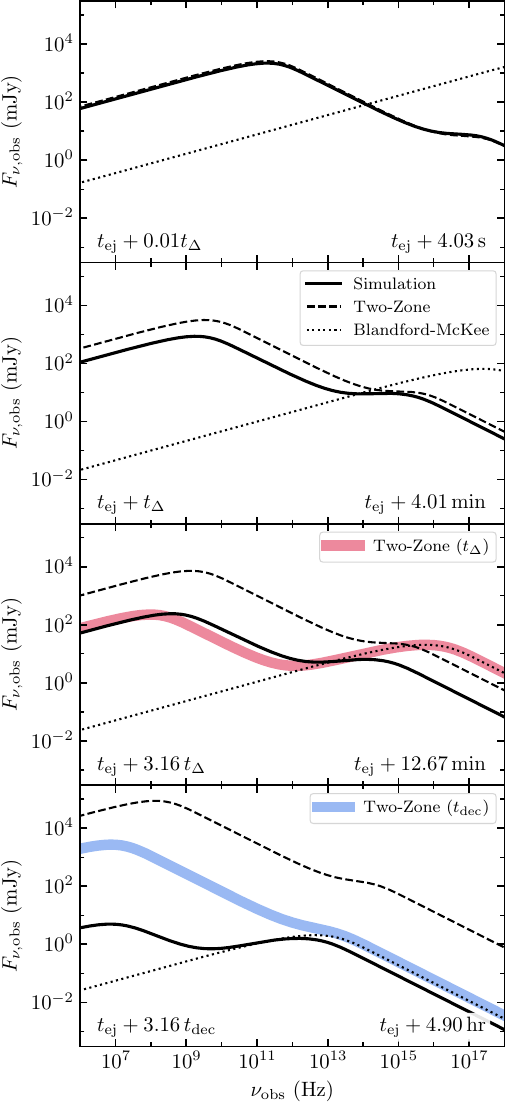}
\caption{Spectral flux density of a mock GRB afterglow with a Newtonian reverse shock expanding into a wind-like ($k=2$) circumburst medium, with $f_{\rm ej}=10^5$, $\gamma_{\rm ej}=31.6$, and $\Delta_{\rm ej}=0.01\times r_{\rm ej}=10^{12}\,{\rm cm}$. The simulation, two-zone, and hybrid spectra include the combined forward- and reverse-shock emission; the standalone Blandford-McKee curve contains only the forward-shock contribution. Spectra from the SRHD simulation are compared to the two-zone model, the Blandford-McKee solution, and the two hybrid prescriptions defined in Sec.~\ref{subsec:hybrid-prescription}, which transition at $t_{\Delta}$ and $t_{\rm dec}$, respectively. The two-zone model agrees well at early times, but discrepancies appear by $t_{\Delta}$ as radial structure develops. Switching at $t_\Delta$ overpredicts the forward-shock emission, while extending the two-zone model to $t_{\rm dec}$ overpredicts the reverse-shocked-ejecta contribution.}
\label{fig:spectral_flux}
\end{center}
\end{figure}

For the synchrotron post-processing, we adopt fixed microphysical parameters $\epsilon_B = 0.01$, $\epsilon_e = 0.1$, and electron power-law index $p=2.23$ \citep[e.g.][]{2013ApJ...776..119L}, and place the source at a distance $d=10^{28}\,\mathrm{cm}$ ($\sim 3\,\mathrm{Gpc}$). Fig.~\ref{fig:spectral_flux} shows spectra at several characteristic times, representing the total emission from the forward- and reverse-shocked regions. We compare spectra computed directly from the SRHD simulation to those obtained from the two-zone model, the Blandford-McKee solution, and the two hybrid prescriptions defined in Sec.~\ref{subsec:hybrid-prescription}, which transition away from the two-zone model at either $t_{\Delta}$ or $t_{\rm dec}$.

These two prescriptions produce distinct, frequency-dependent errors during the interval $t_{\Delta} < t < t_{\mathrm{dec}}$: extending the two-zone model to $t_{\mathrm{dec}}$ most strongly overpredicts the radio through ultraviolet emission, while imposing Blandford-McKee evolution at $t_{\Delta}$ produces the largest discrepancy in the X-ray band. This frequency dependence reflects the different characteristic synchrotron frequencies of the two shocked regions; in the NRS regime, the reverse shock deposits comparatively little thermal energy per particle in the ejecta, resulting in a lower characteristic electron Lorentz factor and synchrotron frequency than in the forward-shocked circumburst medium. The reverse-shock contribution is therefore most important at lower frequencies for the parameters considered here, whereas the more strongly heated forward-shock electrons contribute preferentially at higher frequencies. Consequently, excess thermal energy in the reverse-shocked ejecta primarily enhances the radio through ultraviolet emission in our representative model, while prematurely increasing the forward-shock energy scale is most apparent in the X-ray band. In the earliest snapshot, $(t - t_{\mathrm{ej}}) \ll t_{\Delta}$, the two-zone spectrum agrees closely with the simulation since the shocked regions are still thin and exhibit minimal radial variation of the hydrodynamic quantities. By $t_{\Delta}$, discrepancies have already emerged due to the development of internal structure in the simulation that is absent from the two-zone treatment.

\begin{figure}
\begin{center}
\includegraphics{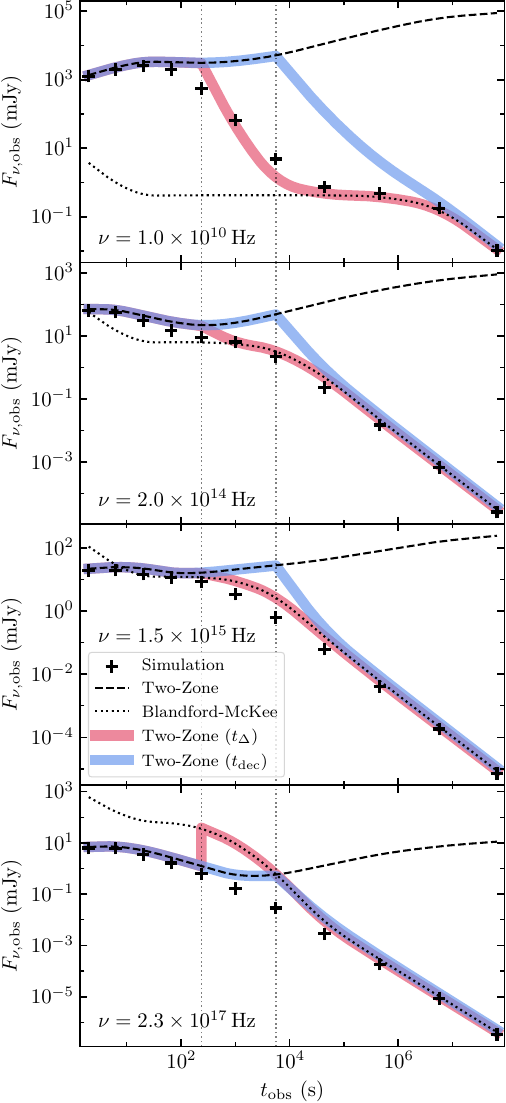}
\caption{Observer-frame light curves for the same mock GRB afterglow shown in Fig.~\ref{fig:spectral_flux}, evaluated at representative radio, near-infrared, ultraviolet, and X-ray frequencies. As in Fig.~\ref{fig:spectral_flux}, each light curve shows the combined forward- and reverse-shock emission, except the Blandford-McKee curve, which only includes a forward-shock component. The vertical dotted lines mark $t_{\Delta}$ and $t_{\mathrm{dec}}$, from left to right. The preferred transition time is frequency-dependent: applying self-similarity early at $t_{\Delta}$ better captures the radio through ultraviolet emission, while extending the two-zone model to $t_{\mathrm{dec}}$ better captures the X-ray emission by avoiding the forward-shock excess introduced by an early Blandford-McKee transition.}
\label{fig:light_curves}
\end{center}
\end{figure}

After $t_{\Delta}$, the two transition choices fail in complementary ways. Switching immediately to the late-time prescription places the forward-shocked circumburst medium on the Blandford-McKee energy scale before enough of the ejecta energy has been transferred into the swept-up external medium. The result is an artificially bright and spectrally hardened forward-shock component, which is most apparent at X-ray frequencies in Fig.~\ref{fig:light_curves}. Extending the two-zone model until $t_{\mathrm{dec}}$ avoids this early forward-shock excess, but keeps the reverse-shocked ejecta artificially hot. Once the reverse shock has crossed the shell, there is no longer unshocked ejecta feeding energy into region 3; continuing to enforce two-zone pressure equilibrium therefore overpredicts the reverse-shock contribution, most visibly in the radio, near-infrared, and ultraviolet bands.

The observer-frame light curves in Fig.~\ref{fig:light_curves} show that these discrepancies are also frequency dependent. In the radio, near-infrared, and ultraviolet bands, extending the two-zone model to $t_{\mathrm{dec}}$ generally overpredicts the transition-phase emission, while imposing Blandford-McKee early gives better agreement with the simulation. In these bands, the dominant error in the delayed-transition prescription is the excess emission from the reverse-shocked ejecta. In the X-ray band, however, the premature transition to Blandford-McKee produces a larger discrepancy over much of the interval $t_{\Delta} < t < t_{\mathrm{dec}}$, since the emission is more sensitive to the artificially enhanced forward-shock component. It is possible to re-normalize the Blandford-McKee energy scale to avoid this jump in flux, but doing so would result in an emission (and energy) deficit at late times $t > t_{\mathrm{dec}}$.

Thus, while the RRS regime is immune to this since $t_{\Delta} \sim t_{\mathrm{dec}}$, there is no single transition time that correctly captures both shocked regions in the NRS regime. Imposing self-similarity prematurely better represents the post-crossing evolution of the reverse-shocked ejecta, but over-predicts the energy in the forward-shocked region during the transition window. Extending the two-zone model beyond $t_{\Delta}$ better represents the forward-shock energy scale near deceleration, but artificially injects excess energy into the reverse-shocked region. These systematic errors propagate from the hydrodynamics to the mock observables, resulting in significant discrepancies from the spectra post-processed from simulation output.

The discrepancies shown here should not be interpreted as implying that either hybrid prescription could never be made to fit observed data. In a full afterglow fit, independent values of $\epsilon_e$ and $\epsilon_B$ in the forward- and reverse-shocked regions could absorb part of the flux and spectral-offset differences shown in Figs.~\ref{fig:spectral_flux} and \ref{fig:light_curves}. However, that compensation would not correct the underlying hydrodynamic error; it would instead transfer the error into the inferred microphysical parameters.

\section{Discussion}
\label{sec:discussion}

The practical consequence of the results presented in the prior section is that the inferred ejecta and microphysical parameters can become sensitive to the assumed hydrodynamic prescription between reverse-shock crossing and relaxation into self-similarity. In the NRS regime, the two common transition time choices misallocate thermal energy between the forward- and reverse-shocked regions, and their discrepancies appear directly in the spectra and light curves when a common set of microphysical parameters is adopted. The same data could therefore favor different ejecta or microphysical parameters depending on whether the model extends the two-zone prescription to $t_{\mathrm{dec}}$ or switches to late-time self-similarity early at $t_{\Delta}$.

For example, if the two-zone model is extended to $t_{\mathrm{dec}}$, the reverse-shocked ejecta remain artificially hot after the reverse shock has crossed the shell. An observational fit would then tend to compensate for this excess reverse-shock emission by lowering the $\epsilon_{e,\mathrm{RS}}$, $\epsilon_{B,\mathrm{RS}}$, or other parameters that reduce the reverse-shock flux. Conversely, if Blandford-McKee evolution is imposed at $t_{\Delta}$, the forward-shocked circumburst medium is placed on too large an energy scale during the transition window. Fits that include X-ray data would then need to compensate by lowering the forward-shock microphysical parameters, or lowering the total energy scale and \textit{increasing} the relevant microphysical parameters to ensure agreement for $t \gtrsim t_{\mathrm{dec}}$.

This concern is most relevant when a Newtonian or thin-shell reverse shock is inferred, suggested, or allowed by the data. Such reverse shocks have been invoked for GRB 130427A \citep{2013ApJ...776..119L} to explain the radio/millimeter and early ($\sim$ hours) UV/optical/NIR emission, and for GRB 160509A \citep{2016ApJ...833...88L} to model the radio emission over the first $\lesssim 10$ days of spectral evolution. Similar reverse-shock interpretations have also been considered (if not necessarily favored) for bursts such as GRB 160625B \citep{2017ApJ...848...69A} and GRB 221009A \citep{2023ApJ...946L..23L}. Additionally, as early-time follow-up and broadband monitoring continue to improve, additional examples of Newtonian reverse-shock emission will likely be identified, making these modeling uncertainties increasingly relevant.

The quantitative discrepancies shown in Figs.~\ref{fig:spectral_flux} and \ref{fig:light_curves} also depend on the approximations adopted in the synchrotron post-processing. Equal-arrival-time and angular integration would mix emission produced at different lab-frame times and locations, tending to smooth sharp features associated with the transition between dynamical prescriptions. Changes in $\epsilon_e$, $\epsilon_B$, $p$, or the treatment of electron cooling would shift the characteristic synchrotron frequencies and can therefore alter the observing bands in which the forward- or reverse-shock discrepancies are most apparent. Other radiative and transfer effects can similarly modify the detailed spectral shape, as discussed below for synchrotron self-absorption. These effects limit the significance of the particular flux ratios and frequency ranges shown in Figs.~\ref{fig:spectral_flux} and \ref{fig:light_curves}. They do not, however, remove the underlying differences in density, pressure, Lorentz factor, and thermal-energy content produced by the hydrodynamic prescriptions, which are present before any radiation model is applied (Figs.~\ref{fig:overlay} and \ref{fig:energy_density}). Our robust conclusion is therefore that the choice of hydrodynamic transition prescription can bias the predicted emission and inferred physical parameters; the precise magnitude and spectral location of that bias are dependent on the radiation and microphysical assumptions.

In such scenarios, particularly in the radio/millimeter bands, SSA can also affect the inferred reverse-shock spectrum, particularly when $\nu_a$ lies near the observing band. We have not included SSA in the synthetic spectra presented in Sec. 4, because our goal is to compare how different hydrodynamic prescriptions populate the emitting shocked regions before applying additional low-frequency transfer effects. Including SSA would change the emergent radio spectrum by suppressing optically thick emission and modifying the location of the observed low-frequency peak, but it would act on the densities and thermal energies supplied by the underlying dynamical model. As such, it would not remove the underlying transition-time ambiguity in the hydrodynamic quantities supplied to the emission calculation.

The present calculations assume a spherically symmetric outflow with uniform ejecta properties, whereas GRB ejecta may be radially stratified in energy and Lorentz factor \citep[e.g.][]{1998ApJ...496L...1R,2000ApJ...535L..33S} and possess substantial angular structure \citep[e.g.][]{2002MNRAS.332..945R,2002ApJ...571..876Z}. For angularly structured jets, we expect the principal hydrodynamic result identified here to remain locally applicable while lateral causal communication between different portions of the jet remains weak: each angular element approximately evolves according to its local energy, Lorentz factor, and shell properties, and regions with a Newtonian reverse shock may therefore still exhibit $t_{\Delta} \ll t_{\rm dec}$. However, different angular elements can have different reverse-shock strengths and transition times, and integration over the visible emitting region may smooth or superpose the resulting discrepancies. Radial structure can modify the problem more fundamentally, since variations in the ejecta Lorentz factor or energy with depth can prolong reverse-shock activity and continued energy injection, so that a single shell-crossing time need not characterize the transition. For example, \citet{2026MNRAS.549ag833S} recently developed an analytic forward-reverse-shock model for radially stratified ejecta in which a long-lived, mildly relativistic reverse shock progressively processes slower material, producing an extended phase of energy transfer that can account for observed X-ray plateaus. The quantitative timescales and spectral discrepancies in this work therefore should not be applied directly to arbitrary structured ejecta; nevertheless, the more general conclusion remains relevant: the end of reverse-shock processing and the establishment of forward-shock self-similarity are physically distinct transitions and may not occur simultaneously. Multidimensional calculations and simulations incorporating radial ejecta structure would be required to quantify this effect for more realistic jet configurations.

Recent semi-analytic frameworks have begun to address some of the aforementioned dynamical limitations directly. The radially structured extension of \texttt{jetsimpy} replaces pressure balance with a global energy-conservation prescription and finds that pressure-balance treatments can substantially overestimate the reverse-shock emission in the thin-shell regime \citep{Wang_2025}. This is qualitatively consistent with our finding that maintaining the pressure-balanced two-zone treatment can artificially enhance the thermal energy and emission of the reverse-shocked ejecta, although the detailed dynamical prescriptions differ between the two calculations. \texttt{VegasAfterglow} similarly adopts an energy-conserving treatment during reverse-shock crossing and evolves the shocked ejecta separately after crossing rather than imposing a single instantaneous transition on both shocked regions \citep{WANG2026100490}. Complementarily, the dynamical prescription underlying \texttt{Magglow} includes a finite transition phase that persists beyond reverse-shock crossing, with the forward shock approaching Blandford-McKee evolution only after the influence of the finite-width ejecta has propagated through the blast \citep{10.1093/mnras/staf1923}. These approaches address different aspects of the shortcomings identified here--specifically, energy-conserving prescriptions avoid relying on pressure equilibrium as the global dynamical closure, while an explicit post-crossing transition recognizes that reverse-shock crossing need not coincide with the establishment of forward-shock self-similarity, which is crucial for Newtonian reverse shocks. Nevertheless, reduced thin-shell models do not resolve the radial structure that develops within the shocked regions, motivating continued calibration against resolved relativistic hydrodynamic simulations.

Another possible solution could entail replacing analytic switching prescriptions with interpolation across suites of relativistic hydrodynamic simulations. This strategy is already common in late-time forward-shock afterglow modeling \citep[e.g.][]{2010ApJ...722..235V, 2012ApJ...749...44V}, but extending it to the early afterglow would require simulation libraries that retain both forward- and reverse-shocked material and span the relevant shell parameters, external density profiles, and reverse-shock strengths. Such an approach would be more computationally expensive to construct, but once tabulated, it could provide a fitting framework that preserves the correct transition behavior without requiring a simulation for every proposed parameter set. Magnetization provides another important axis along which such simulation libraries may need to be extended; \cite{2009A&A...494..879M} showed with one-dimensional RMHD simulations that magnetized ejecta can modify or suppress the reverse shock, alter the onset of the forward-shock emission, and nevertheless relax at late times toward the same Blandford-McKee asymptotic behavior once the shell energy has been transferred to the external medium.

A complementary approach would be to develop a relativistic self-similar description of the interaction regime that follows the two-zone phase. \cite{2024ApJ...975L..14C} derived a set of self-similar solutions for ejecta impacting a circumburst medium in the non-relativistic regime. An analogous relativistic solution -- which can be constructed for arbitrary Lorentz factors \citep{2024ApJ...975L..14C} -- could provide a more efficient bridge between the two-zone model and successive stages of evolution. A more phenomenological alternative would be to modify the two-zone prescription itself so that the forward- and reverse-shocked regions are allowed to leave the uniform-shell approximation at different times. For example, the reverse-shocked ejecta could be transitioned to a post-crossing expansion law at $t_{\Delta}$ while the forward-shocked circumburst medium continues to evolve toward its deceleration energy scale. However, such a hybrid treatment would require a physically consistent prescription for the contact discontinuity and boundary conditions between the regions, so it is less clear whether this would remain a controlled semi-analytic model, or resemble something more like an empirical interpolation scheme \citep[e.g.][]{2022MNRAS.513.4887Z}.

Another approach is to numerically evolve the Zone 2 and Zone 3 solutions on independently moving and stretching grids, joined at the contact discontinuity and treating the FS and RS surfaces as outer boundary conditions. Because the solution remains smooth within each shocked region, each zone could in principle be represented as a spectral element, potentially allowing rapid convergence within a modest number of basis functions. We tested this approach and found the spectral-element scheme to be generally unstable despite the smoothness of the intra-zone solution. Stabilizing such schemes could be the subject of a future work. By contrast, independent grids for Zones 2 and 3 with the shocks as boundaries, coupled to a second-order Godunov scheme, has proven to be stable and economical in mildly or sub-relativistic settings, and that approach is used in \cite{2026arXiv260516490A}.

\section{Conclusions}
\label{sec:conclusions}

We have examined how systematic differences between hydrodynamic prescriptions propagate into mock spectra and light curves, discussing how these differences can introduce model-dependent degeneracies when semi-analytic shock models are used to interpret the early afterglows of GRBs with Newtonian reverse shocks. In this regime, the reverse shock crosses the ejecta at a time $t_{\Delta}$ that can be much earlier than the deceleration time $t_{\mathrm{dec}}$, leaving an extended interval in which the blast is no longer accurately described by the two-zone model but has not yet relaxed into the Blandford-McKee self-similar solution. By comparing semi-analytic prescriptions with high-resolution one-dimensional SRHD simulations and post-processed synchrotron spectra, we find the following:

\begin{enumerate}
    \item The reverse-shock crossing time $t_{\Delta}$ and deceleration time $t_{\mathrm{dec}}$ represent distinct physical transitions, as first discussed in \cite{1995ApJ...455L.143S}. The crossing time $t_{\Delta}$ marks the end of energy injection from unshocked ejecta into the reverse-shocked region, while $t_{\mathrm{dec}}$ marks the time at which the forward shock begins to approach the Blandford-McKee regime. These timescales are comparable when the reverse shock is relativistic, but they can be widely separated when the reverse shock is Newtonian, with an intermediate phase that can last $\sim$ hours in observer time.
    \item The two-zone approximation fails before and after $t_{\Delta}$ in different ways; even before $t_{\Delta}$, internal structure develops in the shocked regions that is absent from the uniform-shell treatment. After $t_{\Delta}$, continuing to enforce pressure equilibrium between the forward- and reverse-shocked regions can artificially maintain the thermal energy of the reverse-shocked ejecta, producing excess reverse-shock emission.
    \item Imposing Blandford-McKee evolution at $t_{\Delta}$ also introduces systematic error. Although switching at $t_{\Delta}$ better reflects the fact that the reverse-shocked ejecta are no longer being heated by newly shocked shell material, it places the forward-shocked circumburst medium onto the self-similar energy scale too early. This prematurely enhances the forward-shock emission, especially in bands where the forward shock emission dominates.
    \item These two choices of transition timescale fail in complementary ways. With fixed microphysical parameters, the hydrodynamic errors appear directly as discrepancies in the predicted spectra and light curves. In observational fits that allow independent forward- and reverse-shock values of $\epsilon_e$ and $\epsilon_B$, some of these discrepancies could instead be absorbed into the fitted microphysics, ejecta properties, circumburst density, or blast energy. Successful fits may therefore remain possible, but the inferred parameters can become degenerate with the assumed post-crossing hydrodynamic prescription.
\end{enumerate}

Our results suggest that early-afterglow modeling in the Newtonian reverse-shock regime should treat the interval between $t_{\Delta}$ and $t_{\mathrm{dec}}$ with care. Extending the two-zone model beyond its physical range of validity can artificially enhance the reverse-shock emission, which in the representative model considered here is most apparent from radio through ultraviolet frequencies, while imposing self-similarity before the forward shock has relaxed to the Blandford-McKee regime can artificially enhance the forward-shock emission, most prominently in the X-ray band for our adopted parameters. Even when such prescriptions reproduce observed light curves, the inferred parameters may reflect compensatory adjustments to the assumed dynamics rather than a unique physical solution. Improving inference in this regime will therefore require dynamical prescriptions that more accurately describe the post-reverse-shock-crossing, pre-self-similar evolution of both shocked regions.

\section*{Acknowledgements}

We thank the anonymous referee for their useful and constructive comments. B.A. and E.R.C. acknowledge support from NASA through the Astrophysics Theory Program, grant 80NSSC24K0897, and through Chandra Award Number 25700383 issued by the Chandra X-ray Observatory Center, which is operated by the Smithsonian Astrophysical Observatory for and on behalf of the National Aeronautics Space Administration under contract NAS8-03060. J.Z. acknowledges support from the National Science Foundation under Grant Number AST-2408034.

B.A. led the project as part of his doctoral thesis research at Clemson University, including the analytic development, numerical simulations, synchrotron post-processing, data analysis, visualization, and manuscript preparation. J.Z. supervised the thesis work at Clemson University and contributed to the original conception of the project, interpretation of the hydrodynamic results, and manuscript revision. E.R.C. contributed to the development of the project idea, interpretation of the results, and manuscript revision. All authors discussed the results and approved the final manuscript.

\section*{Data Availability}

The data underlying this article will be shared on reasonable request to the corresponding author.

\bibliographystyle{mnras}
\bibliography{refs}

\end{document}